\documentclass[prb,amssymb,twocolumn,superscriptaddress]{revtex4}
\usepackage{amsmath}
\usepackage{amssymb}
\usepackage{xcolor}
\usepackage{subfigure}
\usepackage{graphicx}

\newcommand{\be}{\begin{equation}}
\newcommand{\ee}{\end{equation}}

\newcommand{\lf}{\left}
\newcommand{\rg}{\right}
\newcommand{\ra}{\rangle}
\newcommand{\la}{\langle}

\newcommand{\bea}{\begin{eqnarray}}
\newcommand{\eea}{\end{eqnarray}}

\begin{document}

\title{Pauli metallic ground state in Hubbard clusters with Rashba spin-orbit coupling}

\author{Valentina Brosco}
\affiliation{Scuola Internazionale Superiore di Studi Avanzati (SISSA) and 
Democritos National Simulation Center, Consiglio Nazionale delle Ricerche, Istituto Officina dei Materiali (IOM),
Via Bonomea 265, 34136 Trieste, Italy}
\author{Daniele Guerci}
\affiliation{Scuola Internazionale Superiore di Studi Avanzati (SISSA) 
Via Bonomea 265, 34136 Trieste, Italy}
\author{Massimo Capone}
\affiliation{Scuola Internazionale Superiore di Studi Avanzati (SISSA) and 
Democritos National Simulation Center, Consiglio Nazionale delle Ricerche, Istituto Officina dei Materiali (IOM),
Via Bonomea 265, 34136 Trieste, Italy}

\date{\today}

\begin{abstract}
We study the ``phase diagram'' of a Hubbard plaquette with Rashba spin-orbit coupling. We show that the peculiar way in which Rashba coupling breaks the spin-rotational symmetry of the Hubbard model allows a mixing of singlet and triplet components in the ground-state that slows down and it changes the nature of the Mott transition  and of the Mott insulating phases.   
\end{abstract}

\maketitle

Spin-orbit coupling (SOC) refers to the entanglement between the spin and the orbital degrees of freedom of electrons dictated by Dirac equation\cite{WinklerSpinOrbitCoupling}. Affecting the most fundamental symmetries of the Hamiltonian, SOC may give rise to new states of matter \cite{kitaev2001,bansil2016} and open new transport channels.\cite{sinova2015}
Over the years it has  been shown to have profound effects on the phase diagram of correlated insulators,\cite{jackeli2009,witczak-krempa2014} to significantly modify the transport properties of disordered metals,\cite{brosco2016} to change the nature of the superconducting state, \cite{gorkov2001,edelstein2004} just to mention few examples.

The manifestations of SOC in solids and heterostructures  are intimately related to the structure and symmetries of their low-energy Hamiltonian. In bulk oxides with 5d electrons, the main source of spin-orbit coupling is the atomic contribution which acts  ``locally'' modifying the ordering and degeneracy of the atomic orbitals,\cite{witczak-krempa2014} and competes with Hund's exchange coupling 
to determine the electronic properties of the material.\cite{kim2008, kim2009, chun2015}  
A somewhat complementary situation arises in weakly correlated materials where SOC yields non-local spin-dependent effects and it induces non-trivial modifications of the band-structure. In these regards a paradigmatic example is  represented by Rashba spin-orbit coupling.\cite{rashba1960,bychkov1984} 

The latter arises in systems where structural inversion symmetry is broken, as it happens in heterostructures or quantum wells,  and it has long been at the focus of intense research efforts,\cite{manchon2015} since, due to its tunability, it holds promises for spintronics\cite{awschalom2007}  and quantum devices applications. \\ 
Recently,  the discovery that  large values of Rashba coupling can be achieved going outside the realm of weakly correlated metals and semiconductors, at the interface between complex oxides, \cite{benshalom2010,caviglia2010} in organic halide perovskites,\cite{kepenekian2015, zhai2017} on the surfaces of antiferromagnetic insulators \cite{generalov2017} and in the bulk\cite{ishizaka2011,landolt2015} and on the surface\cite{eremeev2012,sakano2013} of polar materials  opened up new research avenues for solid-state physics. 
Remarkable examples are  the connection between ferroelectricity and the Rashba interaction in GeTe, \cite{disante2012} 
or  the temperature dependent interplay of Rashba coupling and magnetic interactions found in HoRh$_2$Si$_2$. \cite{generalov2017} 
  In this context, understanding the role of electronic correlation in Rashba-coupled materials has become of crucial relevance. 
What makes this task even more intriguing is the possibility to investigate how  spin-dependent transport\cite{sinova2015} and topological phases \cite{bansil2016}  are affected by the presence of strong electron correlation and, conversely, how the physics of Mott metal-insulator transition and of Mott insulators can change in the presence of relativistic spin-dependent tunneling terms. 
Beside its fundamental relevance, an understanding of this interplay  would help to design new devices that exploit the tunability of Rashba spin-orbit coupling and the high susceptibility of correlated materials.

We consider the simplest model featuring the interplay between Hubbard-like interactions and a Rashba SOC 
\be\label{RH-Ham}
H=-t\sum_{\la ij\ra}c^\dag_i c_j-t_R\sum_{\la ij\ra}c^\dag_i (\vec \alpha_{ij} \times \vec \sigma)_{z}c_j+ U\sum_{i}n_{i\uparrow}n_{i\downarrow}.
\ee
where $\vec \sigma$ is the vector of Pauli matrices, $\vec \sigma=(\sigma_x,\sigma_y,\sigma_z)$,  $c^\dag_i$ and $c_i$ are spinor creation and annihilation operators, $n_{i\sigma}=c^\dag_{i\sigma}c_{i\sigma}$  and we introduced the vector $\vec \alpha_{ij}=(\alpha^x_{ij},\alpha^y_{ij},0)$, with $\alpha_{ij}^\mu=i(\delta_{ij+a_\mu}-\delta_{ij-a_\mu})$  and $a_\mu$ denoting the unitary translation in the $\mu$ direction.

The model depends on three energy scales: the Hubbard on-site interaction, $U$, the standard hopping $t$, and the ``Rashba tunneling amplitude'', $t_R$, quantifying the energy associated with spin-flipping hopping events.   The strong-coupling limit, $U\gg t,t_R$  has been considered in Refs.[\onlinecite{farrell2014a,sun2017,farrell2014}]. There the authors show that $t_R$ yields a generalized Heisenberg model with Dzjaloshinskii-Moriya and compass interactions\cite{jackeli2009}  and they consider superconductivity \cite{farrell2014a} and  the spin-wave spectrum.\cite{sun2017,farrell2014}
In Ref.[\onlinecite{zhang2015}], instead, cluster dynamical mean field theory\cite{maier2005}  is employed to map out the  phase diagram of the model, showing  that spin-orbit coupling favors a metallic phase at weak coupling and discussing various magnetic orders that arise in the insulating regime.

\begin{figure}[t]
\begin{center}
\includegraphics[width=0.48\textwidth]{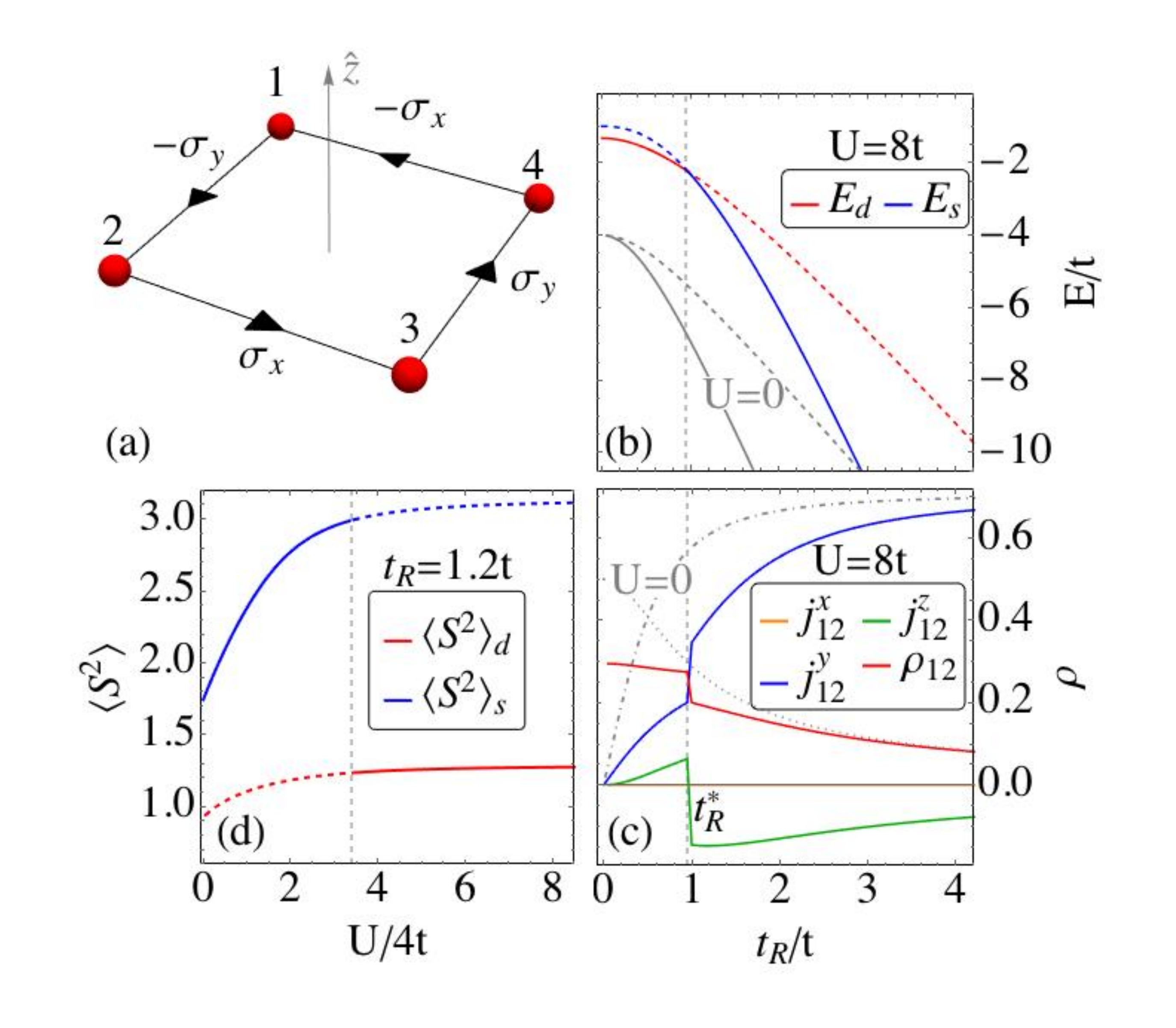}
\caption{(a) Spin-structure of the Rashba spin-orbit coupling Hamiltonian.
(b) Eigenvalue crossing induced by spin-orbit coupling at $U=8t$.  (c) Discontinuity of the bond spin and charge across the transition. (d) Total spin of the plaquette in the states with $s$ and $d$ symmetries.  In all panels dashed and solid lines indicate respectively the ground and excited state energies. Light gray lines represent the non-interacting results in panels (b) and (c).
}
\label{fig:fig1}
\end{center}
\end{figure}

In the present work, in order to gain some analytical understanding, we solve\cite{zitko2011}  a $2\times 2$ Rashba-Hubbard plaquette as depicted in Fig. \ref{fig:fig1}(a). Featuring two spatial directions along which the electrons can hop, this is essentially the minimal system where the chiral nature of Rashba coupling can emerge. \cite{kaplan1983,goth2014}  Furthermore, its Hamiltonian can be diagonalized analytically both at $U=0$ and at $t_R=0$. \cite{noce1996,schumann2002}

We start by discussing  the symmetry properties of the model. Since the Rashba SOC induces a  $SU(2)$ gauge structure on the lattice\cite{guarnaccia2012,tokatly2008}, similarly to what happens in the presence of $U(1)$ gauge fields,\cite{zak1964} the lattice translation group must be defined properly.  In particular, in the case of the plaquette, the presence of SOC implies that all the symmetries of the $D_4$ dihedral group have to be combined with appropriate spin rotations to leave the Hamiltonian invariant. The explicit form of these discrete transformations is, in the case of rotations,
\be\label{eq:gaugesym}
H={\cal U_{\theta}}^{\dag}\,H\,{\cal U_{\theta}} \quad {\rm with} \quad {\cal U_{\theta}}={\cal R}_L(\theta) \otimes e^{-i\frac{\theta}{2} \sigma_z}
\ee  
where ${\cal R}_L(\theta)$ rotates clockwise the whole plaquette state by an angle $\theta = n\pi/2$, with n integer. The single-particle eigenstates of the non-interacting Hamiltonian can be thus classified according to the corresponding quantum numbers and they show a two-fold degeneracy because of the time-reversal symmetry.\cite{supplementary-ev} This implies, in particular, that in the absence of interaction, {\sl i.e.} for $U=0$, at half-filling the ground-state has $s$-wave symmetry {\sl i.e.} it is invariant under $\pi/2$ rotation of the plaquette.  The first two Kramers degenerate doublets, filled in at half-filling, are indeed formed by states acquiring opposite phases under $\pi/2$ rotations.\cite{supplementary} 
At $U=0$, $s$-wave symmetry of the ground-state can be thus traced back to time-reversal symmetry and, in particular, to Kramers degeneracy.
Interestingly, interactions modify this picture. In fact, in the presence of interaction, at half-filling, all states become intrinsically many-body and, since they have an even number of electrons they do not possess Kramers degeneracy, the symmetry of the ground-state under rotations may thus change and the structure of the spectrum is modified, as we show in the following by direct numerical diagonalization of the Hamiltonian 

The symmetries of the Hamiltonian also constrain the average values of the observables, \cite{supplementary} here we consider the bond-charge, or bond-resolved kinetic energy, 
$\rho_{ij}=\la (c^\dag_{i}c_{j}+{\rm H.c})\ra$ and the spin current $\la j^{\mu}_{ij}\ra=-\la ( i\, c^\dag_{i} \sigma_{\mu}c_{j}+{\rm h.c})\ra$, with $i$ and $j$ indicating different lattice sites and $\mu=x,y,z$.
Rotational symmetry implies that the bond-charge and the $z$-component of the spin-current  are the same for all the  bonds of the plaquette. The value of the $x$ and $y$ components of the spin current instead depend on the orientation of the bond. Bonds directed along $x$ feature a non-zero $\la j^y\ra$ and a vanishing $\la j^x\ra$ while bonds directed along $y$ have a non-zero $\la j^x\ra$ and a vanishing $\la j^y\ra$. The invariance under ${\cal U}_{\pi/2}$ in particular yields $\la j^y_{12}\ra=\la j^x_{23}\ra$ while reflection symmetry implies $\la j^x_{12}\ra=\la j^y_{23}\ra=0$. Eventually, the different components of the current  are related by a continuity relation 

\be t \la j^y_{12}\ra=t_{R}\la \lf( \rho_{12}-j^z_{12}\rg)\ra. \label{eq:continuity}\ee

Let us now discuss the properties of the ground-state.
At $t_R=0$ and finite $U$ and $t$ we recover a Hubbard plaquette, whose ground-state is 
 a spin-singlet with $d$-wave symmetry\cite{noce1996,schumann2002} and in the large $U/t$ limit it evolves into a short-range resonating valence bond (RVB) state.\cite{yao2010,scalapino1996}.
   
As we increase the Rashba amplitude $t_R$ a second state having $s$-wave symmetry, therefore more similar to the non-interacting ground-state induced by the Rashba coupling, starts to compete with the $d$-wave RVB-like state. At a certain critical value of $t_R=t_R^*(U)$ the system undergoes a first-order transition from the RVB-like ground-state to the $s$-wave state. The transition is associated to a crossing of energy levels, as shown in Fig. \ref{fig:fig1}(b), and 
it yields a discontinuous behavior in various physical quantities.
As an example, in Fig.\ref{fig:fig1}(c) we show  the different components of the spin current $\la j^{\mu}\ra$ and of the charge on an arbitrary bond, which basically measures the expectation value of the kinetic energy $\rho_{12}=\la (c^\dag_{1}c_{2}+{\rm H.c})\ra$. Beside the discontinuity at $t_R=t_R^*$, in Figure \ref{fig:fig1}(c) we notice that, as the ratio $t_R/t$ increases,   the current becomes completely spin polarized and the bond charge associated to spin-conserving tunneling events is strongly suppressed. The relation between the two behaviors is controlled by the continuity equation (\ref{eq:continuity}).

Figures  \ref{fig:fig1}(b-c) demonstrate the presence of a first-order transition illustrating its most evident consequences. 
We now discuss the origin of  this  transition and the nature of the two competing  states. To this end we recall that since spin-orbit coupling breaks spin-rotational symmetry, the total spin $S^2=\la\vec S\cdot \vec S\ra$ with $ S_\mu=\sum_{i}c^\dag_i\sigma_\mu c_i$  of the system is not a good quantum number. Both states therefore are a mixture of singlet and triplet components with total spin-projection $S_z=0$. Time-reversal symmetry indeed forbids the mixture of states having $S_z=0 $ with states having $S_z\neq 0$.
The total spin $S^2$ in the $s$- and $d$- wave ground-states is shown in Figure \ref{fig:fig1}(d) as a function of $U$. There we see that the total spin, and thus the weight of the triplet, is much larger in the state with $s$ symmetry, this can be easily understood considering that, due to Pauli principle, forming a triplet with $d$ symmetry  requires the occupation of  states with higher momentum and  higher energy than in the case of $s$-wave symmetry. In Fig.\ref{fig:fig1}(c) we also notice that in both states the weight of the triplet increases with $U$ and it saturates in the large $U$ limit.

This suggests the following qualitative picture.
For a finite value of Rashba spin-orbit coupling and small $U$ the ground state has $s$-symmetry,  as we increase $U$ the system exploits the additional degree of freedom given by the breaking of SU(2) spin symmetry to screen the effect of the Hubbard interaction increasing the weight of high-spin configurations. This novel metallic state exploits Pauli principle to override the effect of the Hubbard repulsion and we label it as a "Pauli metal". Such Pauli-enabled screening happens both in the ground ($s$-wave) and in the excited ($d$-wave) state, it is however more efficient in the $s$-wave ground-state since, as explained above, in this state the triplet component can be much larger. 

\begin{figure}[t!]
\begin{center}
\includegraphics[width=0.48\textwidth]{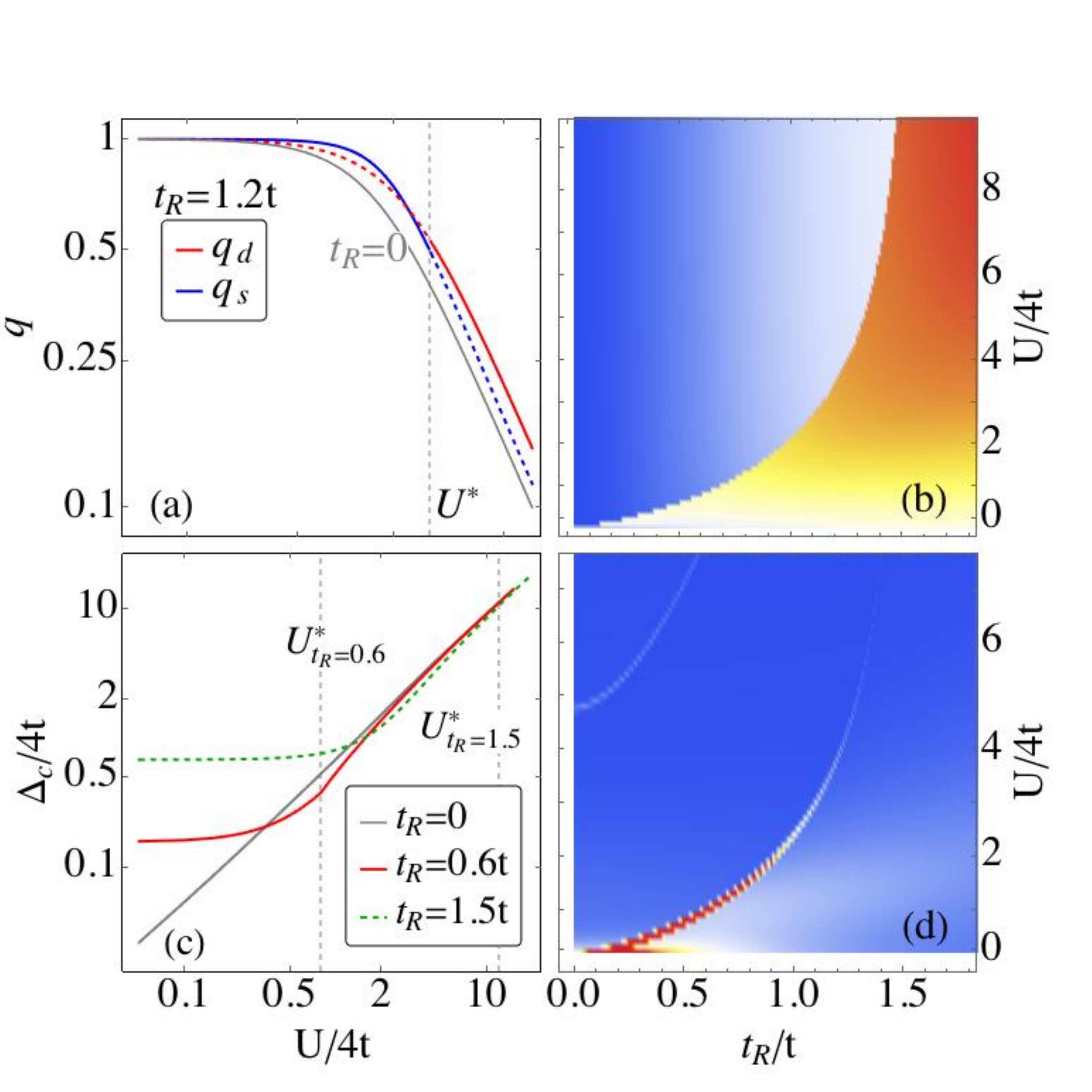}
\caption{
Left panel: Log-log plot of the behavior of the charge gap, $\Delta_c=E_{N+1}+E_{N-1}-2E_{N}$ as a function of $U$ for different values of spin-orbit coupling. Right panel: Absolute value of the  second derivative of the charge-gap with respect to $U$, $|\partial \Delta_c/\partial U^2|$ in the plane $U/t$, $t_R/t$. Ground-state total spin as a function of $U$ and $t_R$. }
\label{fig:fig2}
\end{center}
\end{figure}

The existence of this Pauli-metal has profound consequences on the Mott transition.
A first marker of Mott localization in our small cluster is the kinetic energy reduction factor with respect to the non-interacting value $q=\rho_{12}/\rho_{12}^{(0)}$, which we report in Fig. \ref{fig:fig2}(a) (notice that the U-axis has a logarithmic scale) for the two competing states and, for reference, for the case without Rashba coupling.  The reduction of $q$ measures the correlation-driven localization of the carriers. A Mott transition would correspond to a vanishing $q$ which however can not be realized in our finite-size system, even if a clear crossover takes place between a weak-coupling regime where $q$ changes slowly with $U$ and a strong-coupling regime where $q$ drops faster. 

 The plot clearly shows that the Rashba coupling leads to a larger value of $q$ with respect to $t_R=0$ for both solutions because of  the  Pauli-screening mechanism which increases the metallic character and pushes the Mott transition to larger values of $U$. Interestingly, the $s$-symmetry solution is the least correlated at small $U$ while the $d$-symmetry groundstate is the least correlated for large $U$. Therefore the first-order transition keeps the system in the most metallic of the two states. As we show below, depending on the value of $t_R/t$ Mott localization may or may not coincide with the first-order transition between the two states.

In order to better estimate a critical value $U_c(t_R)$ for the onset of  Mott localization, we consider the charge gap $\Delta_c \equiv E^{N+1}_0+E^{N-1}_0-2 E^{N}_0$ where $E^M_0$ denotes the energy of the $M$-particle ground-state. In  Fig. \ref{fig:fig2}(c) we show  a log-log plot of the charge gap as a function of $U/4t$ for two different values of $t_R$.  As opposed to the case $t_R = 0$, where $\Delta_c$ is linear in $U$ for every value of the interaction, we find a rather well defined crossover. For small values of $U$ when the system is in the $s$-wave state and Pauli screening is effective, the gap is essentially independent on $U$ indicating that it is simply the finite-size gap of our plaquette which would vanish in the thermodynamic limit, while for $U > U^*$ the gap is linear in $U$ as expected in a Mott insulator. This shows that the Pauli-screening qualitatively affects the metallic state and therefore leads to a much better defined Mott transition than in the standard Hubbard model. 

Moreover, for small $t_R$ ($t_R=0.6 t$ in Fig. \ref{fig:fig2}(c)) the localization  transition coincides  with the first-order transition, and the insulating state has $d$-wave symmetry, while for large $t_R$ ($t_R=1.5 t$ in Fig. \ref{fig:fig2}(c)), Mott localization occurs as a crossover within the  $s$-wave symmetry state.  In this case the first-order transition from $s$ to $d$ shifts to extremely large values of $U$. For example for $t_R=1.5\, t$ we find  $U^*> 40 t$.

To illustrate the general structure of the phase diagram in the $\{U,t_R\}$-plane in Fig. \ref{fig:fig2} (d) we show a plot of the absolute value of the second derivative of the charge gap with respect to $U$, {\sl i.e.} $|\partial^2 \Delta_c/\partial U^2|$. Both the first-order transition and  the Mott transition yield a change of slope in the dependence of the gap on $U$ and thus a peak in its second derivative.
However, as one can see in Fig. \ref{fig:fig2}(d), the Mott crossover is not sharp  but it yields a  very broad peak and it appears, in the plot, as a halo  located just below the first-order transition from $s$ to $d$-wave that is instead of first order and thus very sharp. Going to much larger $U$ we notice a very weak but sharp first-order transition line  which signals the onset of Nagaoka's ferromagnetism\cite{nagaoka1965} in the  ground-states with $N\pm1$ electrons.
In Fig. \ref{fig:fig2}(b) we show how the physics we described reveals in the total ground state spin, which obviously has a substantial jump at the first-order transition.

\begin{figure}[b!]
\begin{center}
\includegraphics[width=0.5\textwidth]{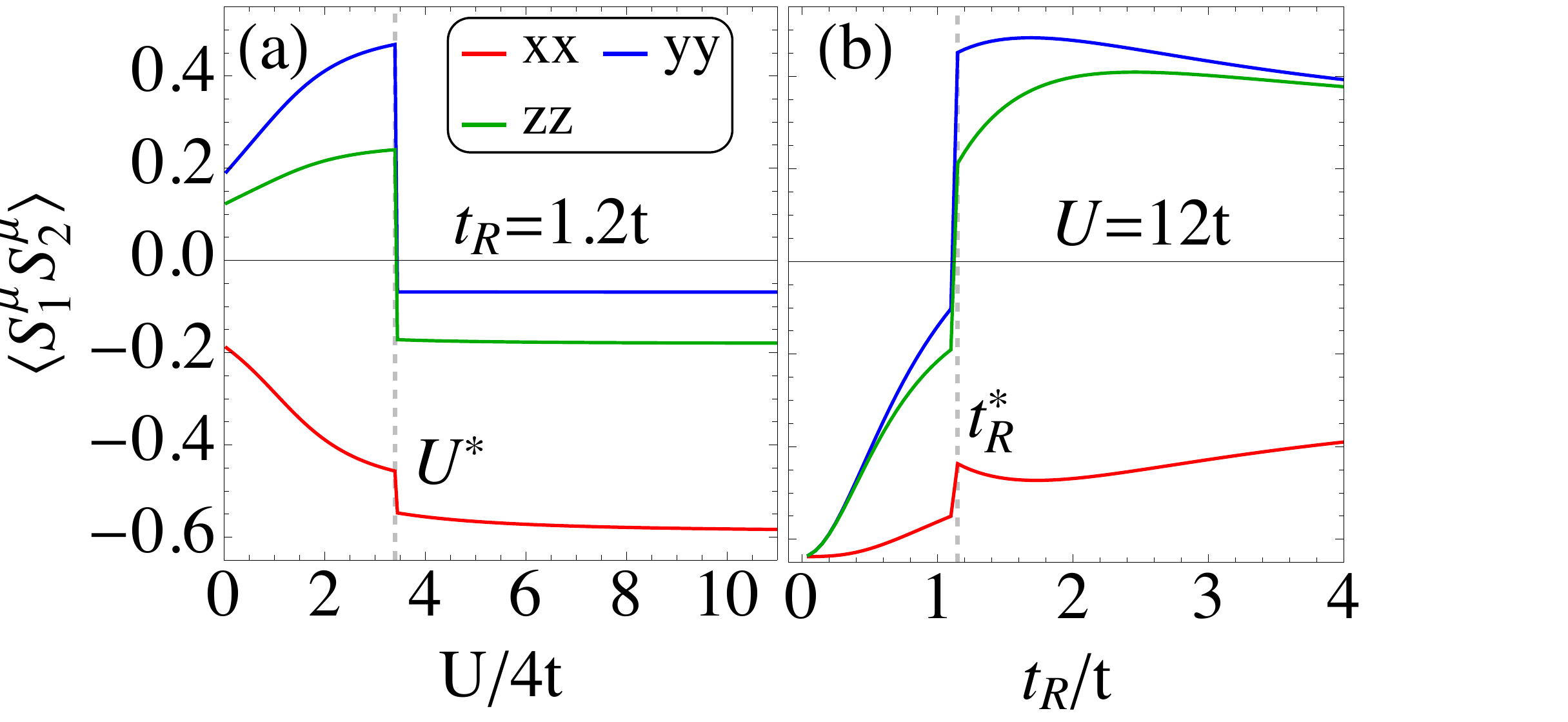}
\caption{Spin-spin correlation functions along the bonds directed along $x$ as a function of $U/t$ (panel a) and of $t_R/t$ (panel b)}
\label{fig:Spinspin}
\end{center}
\end{figure}

 \begin{figure}[t!]
\begin{center}
\includegraphics[width=0.45\textwidth]{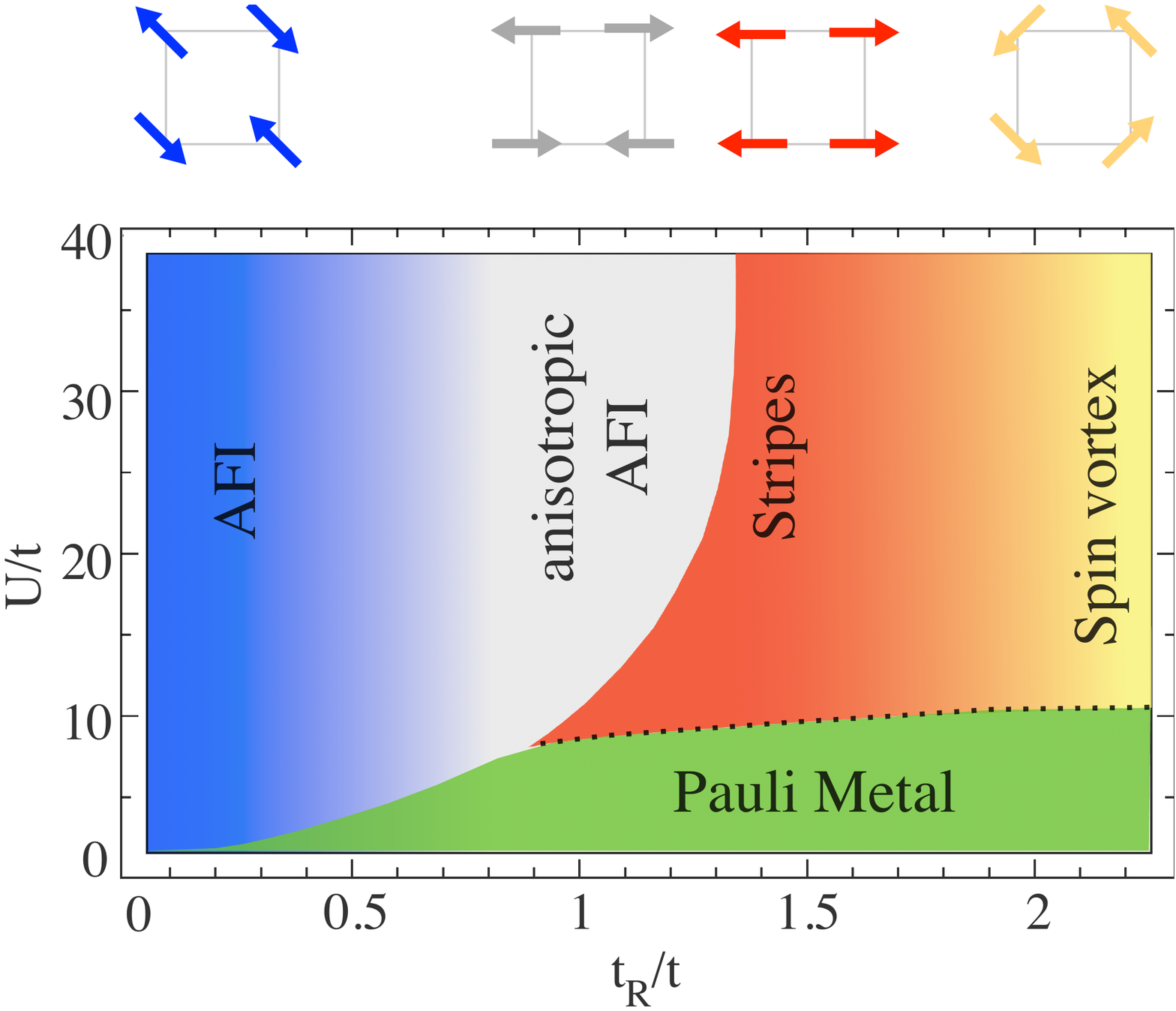}
\caption{Qualitative phase diagram of the Rashba-Hubbard plaquette the in the $(U,t_R)$-plane. The spin patterns characteristic of the four regions are schematically depicted just above the corresponding regimes.}
\label{fig:phasediagram}
\end{center}
\end{figure}

We have shown that depending on the value of $t_R$ the symmetry of the Mott insulating state may be $s$ or $d$, which leads to distinctive magnetic orderings in the different regions of the strong-coupling phase diagram. To characterize the different orderings in Fig. \ref{fig:Spinspin} we show the spin-spin correlation functions, $\la S_i^{\mu}S_j^{\mu}\ra$ along the $1-2$-bond. 
This information is sufficient to reconstruct the spin-spin correlations across the whole plaquette even if, due to the presence of Rashba coupling, the spin-spin correlation functions are anisotropic in spin space and they depend on the direction of the bond. 
However, using the symmetry properties discussed above one can easily show that the spin-spin correlation for orthogonal bonds in the $x-y$ plane may be obtained one from the other by exchanging the $x$ and $y$ components of the spin, so for example $\la S_1^{x}S_2^{x}\ra=\la S_2^{y}S_3^{y}\ra$.

In this figure we can clearly identify two cases.\\
(i) For small $t_R$ 
at large $U$  the ground-state has  $d$-wave symmetry  and, as shown in Fig. \ref{fig:Spinspin}(a), the spin-spin correlations are all negative, implying an anisotropic antiferromagnetic order that reduces to the standard isotropic one as $t_R\rightarrow 0$, as shown in Fig. \ref{fig:Spinspin}(b). 
(ii) For large $U$ and $t_R> t_R^*$ we instead have  an insulating state having $s$-wave symmetry. In this case, on each bond the in-plane spin-spin correlations,  $\la S_i^{x}S_j^{x}\ra$ and $\la S_i^{y}S_j^{y}\ra$ have opposite signs, while the $\la S_i^{z}S_j^{z}\ra$ spin-spin correlations becomes positive, which corresponds, taking into account the symmetries, to a spin-vortex magnetic order for the in-plane spin component and a ferromagnetic order for the $z$ spin-component, in this case reducing   $t_R$ to a value closer to $t_R^*$ distorts the spin-vortex  yielding a striped magnetic-ordering. See the upper panel of Fig. \ref{fig:phasediagram} for a schematic graphical illustration of the different orders.

The above results are summarized in Figure \ref{fig:phasediagram}  where we show a qualitative phase diagram of the plaquette ground-state. For $U<U_c$ and finite $t_R$ we find the Pauli metallic ground-state, at larger $U$ the system undergoes a transition to  a localized state. The localized state may have $s$ or $d$ symmetry depending on the value of $t_R$. In the former case the transition from the Pauli metallic ground-state to the localized state is continuous, as indicated by the dotted line  in Fig.\ref{fig:phasediagram},  while in the latter it becomes of first order. At large $U$ and $t_R<t_R^*(U)$, as we increase $t_R$, we find a continuous transition from an isotropic antiferromagnetic insulator (AFI) to  a more and more anisotropic antiferromagnet. At  $t_R=t_R^*(U)$ the system undergoes a first-order transition to an insulating ground-state having symmetry $s$ and a striped magnetic order. At this point a further increase of the Rashba coupling tends to deform the spin texture of the ground-state yielding a spin vortex magnetic order for $t_R\gg t_R^*$.

In this paper we have solved by a combination of symmetry arguments and exact numerical diagonalization a 2$\times$2 plaquette with Hubbard and Rashba spin-orbit interactions and we have shown that the Rashba coupling promotes a Pauli-screening mechanism which leads to a novel metallic state which is shown to be significantly more robust to Mott localization with respect to the pure Hubbard model. 

The Mott crossover is not the only correlation-driven process of the present model, which shows a first-order transition between a s-wave state stabilized by the spin-orbit coupling and a d-wave state which is closer to the result for the pure Hubbard model. The transition has a profound impact on the properties of the Mott insulator and its magnetic ordering. While the for large $U$ and small $t_R$ the Mott insulator has a standard G-type antiferromagnetic ordering with only  a quantitative anisotropy between the different spin components, for large $t_R$ a phase with a spin-vortex texture is found along the x and y directions while the z component retains antiferromagnetic ordering.

The spirit of this work is to solve exactly and obtain analytical insight, on the minimal cluster where the Rashba coupling can affect the physics of a correlated system, or the basic "chiral unit", which can be used as the building block to construct larger-size Rashba-coupled correlated materials, also in the presence of inhomogeneities as recently found in Refs.[\onlinecite{bindel2016,bovenzi2017}]. Future calculations using variational cluster perturbation theory will help us to elucidate how the physics of the plaquette evolves when the system size grows. The symmetry analysis also helps to rationalize the results of cluster Dynamical Mean-Field Theory calculations\cite{zhang2015}.

We acknowledge useful discussions with A. Amaricci, R. Raimondi and A. Valli and the financial
support from MIUR through the PRIN 2015 program (Prot. 2015C5SEJJ001), SISSA/CNR project ÓSu-
perconductivity, Ferroelectricity and Magnetism in bad metalsÓ (Prot. 232/2015) and European Union under the H2020 Framework Programme, ERC Advanced Grant No. 692670 ÒFIRSTORMÓ.

\bibliographystyle{prsty_no_etal}

\clearpage


%
%

\onecolumngrid
\begin{center}
\large{\textbf{Supplemental Material to ``Pauli metallic ground-state in Hubbard clusters with Rashba spin-orbit coupling''}}

\end{center}
\twocolumngrid
\section*{Symmetries and single-particle spectrum}
\begin{center}
{\setlength\arraycolsep{10pt}
\begin{table*}[]
\begin{tabular}{|l|l|l|}
\hline
\rule{0pt}{5ex} 
\! $E_{0-}=-\varepsilon(1+\cos \alpha)$\,\, &  \begin{tabular}{l}
$V_{0 -+}=e^{-i \pi/4}\sin\alpha/2\, |\pi/2\uparrow\ra +\cos\alpha/2\, |0\downarrow\ra$\\[0.5ex]
$V_{0--}=e^{i \pi/4}\sin\alpha/2\, |-\pi/2\downarrow\ra +\cos\alpha/2\, |0\uparrow\ra$\\[0.1cm]
\end{tabular}& $\Lambda_{0\pm}=e^{\pm i \pi/4}\,\,$\\[0.1cm] 
\hline
\rule{0pt}{5ex} 
\,$E_{1-}=-\varepsilon(1-\cos \alpha)$\,\, &  \begin{tabular}{l}
$V_{1-+}=e^{-i \pi/4}\sin\alpha/2\, |\pi\uparrow\ra +\cos \alpha/2\, |\pi/2\downarrow\ra$\\
$V_{1--}=e^{i \pi/4}\sin \alpha/2\, |\pi\downarrow\ra +\cos \alpha/2\, |-\pi/2\uparrow\ra$\\[0.1cm]
\end{tabular}& $\Lambda_{1\pm}=e^{\pm i 3\pi/4}\,\,$\\[0.1cm]
\hline 
\rule{0pt}{5ex} 
\,$E_{0+}=\varepsilon(1-\cos \alpha)$\,\, &  \begin{tabular}{l}
$V_{0++}=e^{i \frac{\pi}{4}}\cos\alpha/2\, |\pi/2\uparrow\ra -\sin \alpha/2\, |0\downarrow\ra$\\
$V_{0+-}=e^{-i \frac{\pi}{4}}\cos \alpha/2\, |-\pi/2\downarrow\ra -\sin \alpha/2\, |0\uparrow\ra$\\[0.1cm]
\end{tabular}& $\Lambda_{0\pm}=e^{\pm i \pi/4}\,\,$\\[0.1cm]
\hline 
\rule{0pt}{5ex} 
\,$E_{1+}=\varepsilon(1+\cos \alpha)$\,\, &  \begin{tabular}{l}
$V_{1++}=e^{i \pi/4}\cos\alpha/2\, |\pi\uparrow\ra -\sin \alpha/2\, |\pi/2\downarrow\ra$\\
$V_{1+-}=e^{-i \pi/4}\cos \alpha/2\, |\pi\downarrow\ra -\sin \alpha/2\, |-\pi/2\uparrow\ra$\\[0.1cm]\end{tabular}& $\Lambda_{1\pm}=e^{\pm i 3\pi/4}\,\,$\\
\hline
\end{tabular}
\caption{One-particle eigenvalues and eigenvectors of the Hamiltonian H. The third column shows the symmetry properties of the eigenstates.}
\label{table}
\end{table*}}
\end{center}
As explained in the main text, the Rashba-Hubbard Hamiltonian can be written as 
\be\label{RH-Ham}
H= H_0+H_{\rm U}
\ee
where 
\be
\label{non_int}
H_0=-t\sum_{\la ij\ra}c^\dag_i c_j-t_R\sum_{\la ij\ra}c^\dag_i (\vec \alpha_{ij} \times \vec \sigma)_{z}c_j
 \ee
while $H_U$ denotes the Hubbard on-site repulsive interaction
 \be 
 \label{HubbardU}
  H_{\rm U}= U\sum_{i}n_{i\uparrow}n_{i\downarrow}
 \ee
where $\vec \sigma$ is the vector of Pauli matrices, $\vec \sigma=(\sigma_x,\sigma_y,\sigma_z)$,  $c^\dag_i$ and $c_i$ are spinor creation and annihilation operators, $n_{i\sigma}=c^\dag_{i\sigma}c_{i\sigma}$  and we introduced the vector $\vec \alpha_{ij}=(\alpha^x_{ij},\alpha^y_{ij},0)$, with
$\alpha_{ij}^\mu=i(\delta_{ij+a_\mu}-\delta_{ij-a_\mu})$  and $a_\mu$ denoting the unitary translation in the $\mu$ direction.
%
%
We now analyze in more detail the consequences of  the symmetries of the  Hamiltonian $H$ in the case of the plaquette. 

As explained in the main text,  in the presence of SOC the plaquette Hamiltonian is invariant under a group of combined spatial and spin transformations,  below indicated  as $D_{4\sigma}$ that includes four discrete rotations and four reflections affecting both the spatial orientation of the plaquette and the  spin.  

Correspondingly, as we now show,  the single-particle eigenstates can be classified on the basis of their symmetry properties under certain $D_{4\sigma}$ group elements.
Let us focus on rotations, we have 
\be\label{eq:gaugesym}
H={\cal U_{\theta}}^{\dag}\,H\,{\cal U_{\theta}} \quad {\rm with} \quad {\cal U_{\theta}}={\cal R}_L(\theta) \otimes e^{-i\frac{\theta}{2} \sigma_z}
\ee  
where $\theta$ is an integer multiple of $\pi/2$ and ${\cal R}_L(\theta)$ rotates the whole plaquette state by  an angle $\theta$ clockwise.
For $\theta=\pi/2$  transformation ${\cal U}_{\theta}$ has four distinct eigenvalues, $\Lambda_{\lambda\nu}=e^{i \lambda(\pi/4+\nu\, \pi/2)}$ with  $\nu=0,1$ and $\lambda=\pm1$.
The single-particle eigenvectors of the Hamiltonian $H$, $V_{\nu\eta\lambda}$, can be classified using the indices  $\lambda$ and $\nu$ plus a third quantum number, $\eta =0,1$, while the eigenvalues  $E_{\eta\nu}$ do not depend on $\lambda$. Their values are reported in  Table I.
Note that while  $\lambda$ and $\nu$  characterize the symmetry properties of the states, $\eta$ is not directly related to $D_{4\sigma}$. 
In Table I we set $\varepsilon=\sqrt{t^2+2t_R^2}$ and $\alpha=\arctan \sqrt{2}t_R/t$ and we introduced the momentum eigenstates, $| k\sigma\ra=1/2 \sum e^{i k {\rm R}_J}| {\rm R}_J\sigma\ra$, with $| {\rm R}_J\sigma\ra$ denoting a state with one electron on site $J \in[1, \ldots, 4]$ and spin $\sigma$.
%
The two-fold degeneracy of the single-particle spectrum can be also ascribed to  time-reversal symmetry, indeed using the standard representation of the time-reversal operator, ${\cal T}=\sigma_y {\cal K}$ where ${\cal K}$ denotes complex-conjugation,  one can easily show  that the vectors $V_{{\rm \nu}\eta\pm}$ are time-reversed  doublets, {\sl i.e.} ${\cal T}V_{{\rm \nu}\eta+}=V_{{\rm \nu}\eta-}$. 
By looking at Table \ref{table} we can understand that in the absence of interaction, {\sl i.e.} for $U=0$, at half-filling the ground-state has $s$-wave symmetry {\sl i.e.} it is invariant under $\pi/2$ rotation of the plaquette. 
The presence of Kramers degeneracy indeed implies that the non-interacting four electron ground-state is constructed by filling  the first two doublets and thus it involves pairs of states that acquire opposite phases under $\pi/2$ rotations. 
The $s$-wave symmetry of the non-interacting ground-state can be thus traced back to Kramers degeneracy and it can be ultimately demonstrated in general terms using the following three relations:
\be 
[{\cal T},\, H]=0, \,\,  [{\cal U}_\theta,\, H]=0 \,\, {\rm and}\,\, {\cal T} {\cal U}_\theta {\cal T}={\cal U}_{-\theta}
\ee
analogously to what  proposed  in Ref. \cite{yao2010}.
%

\section*{Continuity equations}
\label{app-CE}
Starting from the Hamiltonian $H$  (Eq.\eqref{RH-Ham}) we can write the following Heisenberg equation of motion for the local spin  density $S_{i}^\mu$:
\be
\label{eq:EvSpin1}
i \partial_t S_{i}^\mu=-i t \,{\rm div}\lf[ j^{\mu}_{i}\rg]-t_R\sum_j \lf[c^\dag_i\sigma^\mu(\vec \alpha_{ij} \times \vec \sigma )_{z}c_j-{\rm H.c.}\rg]
\ee
where $S_{i}^\mu=c^\dag_{i}\sigma^\mu c_{i}$ and ${\rm div}\lf[ j^{\mu}_{i}\rg]=\sum_{\kappa}(j^\mu_{i,i+a_{\kappa}}-j^\mu_{i,i-a_{\kappa}})$. Let us consider the case of a non homogeneous where $\alpha_{ij}^{\mu}=i(\delta_{i,j+a_{\mu}}\gamma^{\mu}_{i,i-a_{\mu}}-\delta_{i,j-a_{\mu}}\gamma^{\mu}_{i,i+a_{\mu}})$, and $\gamma^{\mu}_{ij}=\gamma^{\mu}_{ji}$. Equation (\ref{eq:EvSpin1}) reads 
\bea
\label{eq:EvSpin2}
i\partial_{t}S_{i}^{x}&=&-it\text{div}j_{i}^{x}-it_{R}\Big[\gamma_{i,i+y}^{y}\rho_{i,i+y}-\gamma_{i,i-y}^{y}\rho_{i,i-y}\nonumber \\
 &+&\gamma_{i,i+x}^{x}j_{i,i+x}^{z}+\gamma_{i-x,i}^{x}j_{i-x,i}^{z}\Big],\nonumber \\
i\partial_{t}S_{i}^{y}&=&-it\text{div}j_{i}^{y}-it_{R}\Big[\gamma_{i,i-x}^{x}\rho_{i,i-x}\\
 &-&\gamma_{i,i+x}^{x}\rho_{i,i+x}+\gamma_{i,i+y}^{y}j_{i,i+y}^{z}+\gamma_{i,i-y}^{y}j_{i-y,i}^{z}\Big],\nonumber  \\
i\partial_{t}S_{i}^{z}&=&-it\text{div}j_{i}^{z}+it_{R}\Big[\gamma_{i-y,i}^{y}j_{i-y,i}^{y}\nonumber \\
 &+&\gamma_{i-x,i}^{x}j_{i-x,i}^{x}+\gamma_{i,i+y}^{y}j_{i,i+y}^{y}+\gamma_{i,i+x}^{x}j_{i,i+x}^{x}\Big].\nonumber 
\eea
We notice that the previous set of equations agrees with those obtained in previous works\cite{seibold2015,seibold2017}.
The geometry of the plaquette implies: $\gamma_{12}^{x}=1,\,\gamma_{23}^{y}=1,\,\gamma_{34}^{x}=1$ and $\gamma_{41}^{y}=1$.
Eq. (\ref{eq:EvSpin2}) applied to site ``1'' of the plaquette in turn yields
\bea
\label{eq:EvSpin3}
\dot S^x_{1}&=&-t \lf(j^x_{12}-j^x_{41}\rg)- t_{R}\lf( j^z_{12}-\rho_{14}\rg)\label{sx1}\nonumber\\
\dot S^y_{1}&=&-t\lf( j^y_{12}-j^y_{41}\rg)- t_{R}\lf( j^z_{41}-\rho_{12}\rg)\label{sy1}\\
\dot S^z_{1}&=&-t\lf(j^z_{12}- j^z_{41}\rg)+t_R\lf(j^x_{12}+j^y_{41}\rg).\label{sz1}\nonumber \eea

As we show below the symmetries of the Hamiltonian constrain the average  values of the spin and charge currents so that in the ground-state (i) the  current $j^{z}$  and the bond-charge are  homogeneous across the plaquette, 
\be
j_{12}^{z}=j_{23}^{z}=\ldots \quad {\rm and }\quad \rho_{12}^{z}=\rho_{23}^{z}=\ldots \label{c1}
\ee
while  $j^{x}$ and  $j^{y}$ satisfy the following relations:
\be \la j^y_{12}\ra=\la j^x_{41}\ra\quad {\rm and }\quad\la j^x_{12}\ra=\la j^y_{34}\ra=0.\label{c2}\ee 
Averaging Eqs. (\ref{eq:EvSpin3})  over the ground-state, imposing  $\la \dot S^\mu_{i}\ra=0$ and using the constraints, Eqs.(\ref{c1}-\ref{c2}),  we eventually obtain the  continuity equation introduced in the main text
\be
0=t_{R}\la \lf( \rho_{12}-j^z_{12}\rg)\ra-t \la j^y_{12}\ra.  \label{c-E}
\ee
%
%
%


To conclude this section we show how the constraints, Eqs.(\ref{c1}-\ref{c2}), on the currents can be obtained starting from the  invariance of Hamiltonian (\ref{RH-Ham}) under $D_{4\sigma}$ transformations, we have:
\bea
\label{eq:symmetries1}
\langle j^{x}_{12}\rangle&=&\langle {\cal U^{\dagger}}_{\pi/2}j^{x}_{12}{\cal U}_{\pi/2}\rangle=-\langle j^{y}_{23}\rangle,\nonumber\\
\langle j^{y}_{12}\rangle&=&\langle {\cal U^{\dagger}}_{\pi/2}j^{y}_{12}{\cal U}_{\pi/2}\rangle=\langle j^{x}_{23}\rangle,\\
\langle j^{z}_{12}\rangle&=&\langle {\cal U^{\dagger}}_{\pi/2}j^{z}_{12}{\cal U}_{\pi/2}\rangle=\langle j^{z}_{23}\rangle,\nonumber
\eea
where the unitary transformation ${\cal U^{\dagger}}_{\pi/2}$ is defined in Eq. (\ref{eq:gaugesym}).
From the last equation follows that the z-component of the spin-current is homogeneous across the plaquette. Another element of $D_{4\sigma}$ is ${\cal U}_{r}={\cal R}(1\to2,3\to4) \otimes e^{-i\frac{\pi}{2} \sigma_x}$, where ${\cal R}(1\to2,3\to4)$ is a reflection respect to a vertical axis. By applying the latter transformation to $j^{x}_{12}$
\be
\label{eq:symmetries2}
\langle j^{x}_{12}\rangle=\langle{\cal U}^{\dagger}_{r} j^{x}_{12} {\cal U}_{r}\rangle=\langle j^{x}_{21}\rangle,
\ee 
which implies $\langle j^{x}_{12}\rangle$=0. We conclude that $\langle j^{y}_{23}\rangle=\langle j^{x}_{34}\rangle=\langle j^{y}_{41}\rangle=0$.

\end{document}